# Symbolic Solutions of Simultaneous First-order PDEs in One Unknown

**Célestin Wafo Soh**

We propose and implement an algorithm for solving an overdetermined system of partial differential equations in one unknown. Our approach relies on Bour-Mayer method to determine compatibility conditions via Jacobi-Mayer brackets. We solve compatible systems recursively by imitating what one would do with pen and paper: Solve one equation, substitute the solution into the remaining equations and iterate the process until the equations of the system are exhausted. The method we employ for assessing the consistency of the underlying system differs from the traditional use of differential Gröbner bases yet seems more efficient and straightforward to implement. We are not aware of a computer algebra system that adopts the procedure we advocate in this work.

## ■ Introduction

The search of solutions of many problems leads to overdetermined systems of partial differential equations (PDEs). These problems comprise the computation of discrete symme-tries of differential equations [1], the calculation of differential invariants [2] and the determination of generalized Casimir operators of a finite-dimensional Lie algebra [3]. In this paper, we focus solely on the integration of simultaneous systems of scalar first-order PDEs i.e. our systems will have at least two equations, one dependent variable (unknown function) and several independent variables. Our ultimate goal is to automate the search of general symbolic solutions of these systems. The approach we adopt uses Bour-Mayer method [4] to find compatibility conditions (i.e. obstructions to the integrability) of the underlying system of PDEs and iteratively prepend these compatibilities conditions to it until a consistent or an inconsistent system is found. Note that such an approach differs from the traditional approach which uses differential Gröbner bases [5] to discover compatibility conditions. When it is applicable, it also has several advantages over the differential Gröbner basis approach which include ease of implementation and efficiency. Recently, using differential geometric machinery, Bour-Mayer method has been extended by Kruglikov and Lychagin [6] to systems of PDEs in several dependent, independent variables and mixed orders (i.e. the orders of individual equations in the system do not have to be the same). In our approach, for the situation where the completion process leads to a consistent system, we solve the latter by imitating what one would do with pen and paper: Solve one equation, substitute it into the next equation and continue the process until the equations of the system are exhausted. In order to fix ideas, consider a system of PDEs

$$F_i(x_1, x_2, ..., x_n, z, p_1, p_2, ..., p_n), \quad i = 1:m, \tag{1}$$

where $x_1$ to $x_n$ are the independent variables, $p_k$ is the partial derivative of the unknown function z with respect to $x_k$, and the rank of Jacobian matrix $J = \left[\frac{\partial F_i}{\partial p_j}\right]$ is $m$. The system (1) is integrable i.e. admits a sufficiently smooth solution, provided the expressions of $p_1$ to $p_n$ derived from it satisfies the conditions

$$\frac{dp_i}{dx_j} = \frac{dp_j}{dx_i} \qquad i < j = 1:n. \tag{2}$$

Bour and Mayer (see e.g. [4]) showed that the system (1), subject to the condition on the Jacobian matrix of the $F_i$'s with respect to the $p_j$'s, is integrable if and only if the Jacobi-Mayer

$$[F_i, F_j] := \sum_{k=1}^{n} \frac{\partial F_i}{\partial x_k}\frac{\partial F_j}{\partial p_k} - \frac{\partial F_j}{\partial x_k}\frac{\partial F_i}{\partial p_k} = 0, \quad i < j = 1:m, \tag{3}$$

whenever the system (1) is satisfied. From now on, the phrase `` $[F_i, F_j] = 0$ whenever the system (1) is satisfied will be abbreviated as $[F_i, F_j]|_{(1)} = 0$.

For a given system (1) satisfying the nondegeneracy condition mentioned above, four cases arise. The first case is when $m = n$ and all the Jacobi-Mayer brackets vanish whenever the system (1) is satisfied. In this case we can solve the system (1) for $p_1$ to $p_n$. The solution of the system is then obtained by integrating the exact differential form $dz - \sum_{i=1}^{n} p_i \, dx_i$. The second case is when there are distinct indices $a$ and $b$ such that $[F_a, F_b]|_{(1)} = \phi(x_1, x_2, ..., x_n, z) \neq 0$. In such instance, the system (1) is incompatible and there are no solutions. In the third case, $m < n$, and all the Jacobi-Mayer brackets vanish on the system (1). We must supplement the system (1) with additional equations until we find ourselves either in the first case or in the second case. These equations are obtained by solving the system of linear first-order PDEs $[F_\lambda, F_\mu] = 0$, where $\lambda = 1:n$ and $\mu = m+1:n$. For example the additional equation $F_{m+1} = a_1$, where $a_1$ is an arbitrary constant, is begotten by solving the system of linear first-order PDEs $[F_i, F_{m+1}] = 0$, where $i = 1:m$. The solution of the completed system depends on $m - n + 1$ arbitrary constants. We obtained the general solution of the initial system (1), by expressing one of the arbitrary constants as a function of the remaining ones, and then eliminating the remaining constant between the resulting equation and their first-order partial derivatives with respect to the arbitrary constants. The final case i.e. the fourth one corresponds to the eventuality, where some brackets are zero on the system (1), and other brackets have the form $[F_a, F_b]|_{(1)} = \psi_{ab}(x_1, x_2, ..., x_n, z, p_1, p_2, ..., p_n)$, where the $\psi_{ab}$'s depend at least on some $p_i$'s. In this case, we must prepend the equations $\psi_{ab}(x_1, x_2, ..., x_n, z, p_1, p_2, ..., p_n) = 0$ to the system (1), and proceed as in the third case. The procedure we have just described is the essence of Bour-Mayer approach to the solution of the system (1). Note that in such approach, one has to solve overdetermined systems of linear scalar PDEs, and ensure that the equations one add to the initial system are compatible with them, and the equations of resulting systems are linearly independents. In our implementation of Bour-Mayer approach, we complete the initial system (1) by prepending to it the appropriate compatibility constraints prescribed by Jacobi-Mayer brackets, until we obtain either a compatible system or an incompatible one. We iteratively solve the compatible system obtained, starting from compatibility constraints, and using the build-in function **DSolve**. The remainder of this paper is devoted to the implementation and test of our approach.

# Implementation and Tests

Here we focus on the coding of the algorithm described in the introduction. Specifically, we shall start by iteratively solving a system of consistent first-order PDEs in one depen-dent variable. Then, we shall implement the test of consistency of a system of first-order PDEs in one unknown. Finally we shall couple the last two codes in such a way that a single function is used for computing the general solution of the input system when it exists or indicates whether it is inconsistent.

## Iterative solution of a consistent system of first-order PDEs in a single unknown

Our code for implementing the iterative solution of a compatible system of scalar first-order PDEs is made of a main function (**solveCompatiblePdes**) and three helper functions (**pdesToRules**, **rulesToSolution** and **exprToFunction**). The function **pdeToRules** is a recursive function which consumes as input the system to be solved (**syst**), the dependent variable (**unKnown**), the list of independent variables (**indepVars**), a container for the list of successive solutions (**sol**), a list of equations that could not be solved (**unsolvedEqs**), a string (**symb**) that will be used as root to form the names of intermediate dependent variables and a variable (**count**) that is used to count and name intermediate dependent variables. Its output is made of a list of rules and a list of unsolved equations. The function **pdeToRules** mimics what one would to by hand when solving a system of first-order PDEs in one unknown: Solve an equation and substitute its solution into remaining equations until the equations of the system are exhausted. At each stage, the number of independent variables is reduced by one and one must rename the variables before proceeding. Another issue that arises in the implementation is that the dependent variables are curried functions that must be uncurried to insured that the chain rule is applied properly during substitution into the remaining PDEs. Perhaps, this is the trickiest part of our implementation. The function **rulesToSolution** takes the output of **pdeToRules** and convert it into the solution of the system to be solved. The helper function **exprToFuction** converts an expression (**expr**) depending on several variables (**vars**) into a pure function of these variables. Finally, the function **solveCompatiblePdes** composes **pdesToRules** and **rulesToSolution** to solve a compatible system of scalar PDEs. Its inputs are similar to those of **pdesToRules** and its outputs are formatted as those of **rulesToSolution**.

```
pdesToRules[syst_, unKnown_, indepVars_, sol_: {},
   unsolvedEqs_: {}, symb_: "x", count_: 0] :=
  Module[{currentSol, newIndepVars = {}, newSyst,
    nameUnknown = "", temp, newUnsolvedEqs, f},
   If[ syst == {}, {sol, unsolvedEqs},
```

```
      If[First@syst === False, {{}, unsolvedEqs},
       If[
        Head[DSolveValue[First@syst, unKnown, indepVars]] ===
         DSolveValue, pdesToRules[Rest[syst], unKnown,
         indepVars, sol, Append[unsolvedEqs, First@syst],
         symb, count + 1],
        currentSol = DSolveValue[First@syst, unKnown,
          indepVars, GeneratedParameters → (C[# + count] &)];
        f = # /. {C[z_][t__][y__] ⧴ C[z] @@ {t, y}} &;
        FixedPoint[f, currentSol @@ indepVars] /.
         C[_][z__] ⧴ (newIndepVars = {z});
        currentSol @@ indepVars /.
         { C[z_] ⧴ ( nameUnknown = C[z]) };
        currentSol = FixedPoint[f, currentSol];
        newSyst = Rest[syst] /. {unKnown -> currentSol};
        newUnsolvedEqs = unsolvedEqs /. unKnown → currentSol;
        newSyst = Select[newSyst, Not[# === True] &];
        If[newIndepVars ≠ {},
         temp =
          First@Solve[# == Unique[symb] & /@ newIndepVars,
            indepVars];
         {newIndepVars, newSyst, newUnsolvedEqs} =
          Map[ Simplify[# /. temp ,
             TransformationFunctions →
              {Automatic, PowerExpand}] &,
            {newIndepVars, newSyst, newUnsolvedEqs}],
         ## & ];
        currentSol = Append[sol, unKnown → currentSol];
        pdesToRules[newSyst, nameUnknown, newIndepVars,
         currentSol, newUnsolvedEqs, symb, count + 1]]]];

exprToFunction[expr_, vars_] :=
  Function@expr /. MapIndexed[#1 → Slot @@ #2 &, vars];

rulesToSolution[lst_, unKnown_, indepVars_] :=
  Module[{s = First@lst, temp1, temp2 , count = 0},
   If[s === {}, {},
```

```
    temp1 = exprToFunction[
       Fold[#1 /. #2 &, Values[First@s] @@ indepVars, Rest@s],
       indepVars]];
    temp2 = First@Rest@lst /. unKnown -> temp1;
    {unKnown @@ indepVars -> temp1 @@ indepVars, temp2}] ;

solveCompatiblePdes[syst_, unKnown_, indepVars_,
    sol_ : {}, constraints_ : {},  symb_ : "x"] :=
  rulesToSolution[pdesToRules[syst, unKnown, indepVars,
      sol, constraints,  symb] , unKnown, indepVars];
```

## ❒ Compatibility test and completion

This part is dedicated to the implementation of the compatibility test provided by Bour-Mayer method as described in the introduction. The function implementing the compatibility test is name **compatibilityQ**. It consumes as inputs the underlying system of PDEs (**syst** ), the dependent variable (**depvar**) and the list of independent variables (**indepVars**). It outputs a list made of two items: the first item tells whether the system is compatible whereas the second entry provides the completed system. There are three additional functions, **mayerBrackets**, **mayerBracketSyst**, and **derivativeQ**, whose contract deals respectively with the computation of Jacobi-Mayer as defined in Eq. (3), the pairwise Jacobi-Mayer brackets of a system restricted to the system, and detecting whether an expression contains a derivative of the unknown function.

```
mayerBrackets[f_, g_, depvar_, indepVars_] :=
  Module[{p = D[depvar @@ indepVars, #] & /@ indepVars,
    func = Function[{x, y}, D[x, #] & /@ y]},
   func[f, p].func[g, indepVars] -
     func[g, p].func[f, indepVars] // Simplify];

mayerBracketsSyst[syst_, depvar_, indepVars_] :=
  Module[
   {p = Solve[# == 0 & /@ syst,
      D[depvar @@ indepVars, #] & /@ indepVars]},
   If[p == {}, {},
    Flatten[
      Table[ Table[mayerBrackets[syst[[i]], syst[[j]],
          depvar, indepVars], {i, 1, j - 1}],
        {j, 1, Length[syst]}]] /. First@p // Simplify]];

derivativeQ[expr_, dep_, indep_] :=
  Module[{temp = 0},
   expr /. D[__][dep] @@ indep :> (temp = temp + 1;);
   temp > 0];

compatibilityQ[syst_, depvar_, indepVars_] :=
  Module[
   {brkts = mayerBracketsSyst[syst, depvar, indepVars],
    temp},
   If[brkts == {}, {False, {}},
    brkts = Select[brkts, Not[# === 0] &];
    If[ brkts == {}, {True, syst},
     temp = Select[brkts,
        ! derivativeQ[#, depvar, indepVars] &];
      If[temp == {} && Length[syst] + Length[brkts] <=
         Length[indepVars],
       compatibilityQ[Join[brkts, syst], depvar, indepVars],
       {False, {}} ] ]]];
```

◻ **Putting everything together**

Here we employ the functions of the previous subsections to solve an overdetermined system of first-order PDEs in one unknown. There are two functions: **solveOverDeterminedScalarFirstOrderPdes** and **solutionQ**. The semantic of the first function is included in its name and it consumes the system to be solve (**syst**), its dependent (**depvar**) and independent variables (**indepVars**). The second function verifies whether a given rule (**sol**) defines a solution of a system system of first-order PDEs (**syst**) in one unknown. We shall demonstrate how to call these functions in the next subsection which deals with tests.

```
solveOverDetScalarFirstOrderPdes[syst_, depvar_,
    indepVars_] :=
  Module[
   {cSyst = compatibilityQ[syst /. a_ ⩵ b_ ⧴ a - b, depvar,
      indepVars]},
   If[First@cSyst, solveCompatiblePdes[
     # ⩵ 0 & /@ First@Rest@cSyst, depvar, indepVars], {}]];

solutionQ[syst_, depvar_, indepVars_, sol_] :=
  SelectFirst[syst,
    Not[
     Simplify[
       # /. depvar → exprToFunction[Values@sol,
         indepVars]] ⩵⩵⩵ True] &] ⩵⩵⩵ Missing["NotFound"];
```

◻ **Tests**

This subsection is chiefly concerned with the test of the function **solveOverDetScalarFirstOrderPdes**. The examples are taken from various sources that we shall specify as we proceed. In all the examples, we have suppressed the warnings, through the build-in function **Quiet**, for convenience.

▪ *Test 1*

The examples we present here arises in the search of differential invariants of hyperbolic PDEs [2].

```
xs = {h, k, ht, hx, kt, kx}; ys = J @@ xs;
syst1 = {h * D[ys, ht] + k * D[ys, kt] == 0,
   h * D[ys, hx] + k * D[ys, kx] == 0,
   h * D[ys, h] + k * D[ys, k] + 2 * ht * D[ys, ht] +
      hx * D[ys, hx] + 2 * kt * D[ys, kt] + kx * D[ys, kx] == 0,
   h * D[ys, h] + k * D[ys, k] + ht * D[ys, ht] + 2 * hx * D[ys, hx] +
      kt * D[ys, kt] + 2 * kx * D[ys, kx] == 0};
sol1 = solveOverDetScalarFirstOrderPdes[syst1, J, xs] //
   Quiet
```

$$\left\{ J[h, k, ht, hx, kt, kx] \to C[4]\left[\frac{k}{h}, \frac{(-ht\,k + h\,kt)\,(-hx\,k + h\,kx)}{h^5}\right], \{\}\right\}$$

```
solutionQ[syst1, J, xs, First@sol1]
```

True

```
xs = {h, k, ht, hx, kt, kx, htt, htx, hxx, ktt, ktx, kxx};
ys = J @@ xs;
eq1 = k * D[ys, kxx] + h * D[ys, hxx] == 0;
eq2 = k * D[ys, ktt] + h * D[ys, htt] == 0;
eq3 = k * D[ys, kx] + h * D[ys, hx] + 3 * kx * D[ys, kxx] +
    kt * D[ys, ktx] + 3 * hx * D[ys, hxx] + ht * D[ys, htx] == 0;
eq4 = k * D[ys, kt] + h * D[ys, ht] + kx * D[ys, ktx] +
    3 * kt * D[ys, ktt] + hx * D[ys, htx] + 3 * ht * D[ys, htt] == 0;
eq5 = kxx * D[ys, kxx] + 2 * ktx * D[ys, ktx] + 3 * ktt * D[ys, ktt] +
    hxx * D[ys, hxx] + 2 * htx * D[ys, htx] + 3 * htt * D[ys, htt] +
    kx * D[ys, kx] + 2 * kt * D[ys, kt] + hx * D[ys, hx] +
    2 * ht * D[ys, ht] + k * D[ys, k] + h * D[ys, h] == 0;
eq6 = 3 * kxx * D[ys, kxx] + 2 * ktx * D[ys, ktx] + ktt * D[ys, ktt] +
    3 * hxx * D[ys, hxx] + 2 * htx * D[ys, htx] + htt * D[ys, htt] +
    2 * kx * D[ys, kx] + kt * D[ys, kt] + 2 * hx * D[ys, hx] +
    ht * D[ys, ht] + k * D[ys, k] + h * D[ys, h] == 0;
syst2 = {eq1, eq2, eq3, eq4, eq5, eq6};
sol2 = solveOverDetScalarFirstOrderPdes[syst2, J, xs] //
  Quiet
```

$$\left\{ J[h, k, ht, hx, kt, kx, htt, htx, hxx, ktt, ktx, kxx] \to \right.$$
$$C[6]\left[\frac{k}{h}, -\frac{-h\,htx + ht\,hx}{h^3}, -\frac{hx\,kt - h\,ktx - \frac{ht\,(hx\,k - h\,kx)}{h}}{h^3}, \right.$$
$$-\frac{h\,(-3\,hx^2\,k + 3\,h\,hx\,kx - h\,(-hxx\,k + h\,kxx))}{(hx\,k - h\,kx)^2},$$
$$\frac{(ht\,k - h\,kt)\,(hx\,k - h\,kx)}{h^5},$$
$$\left.-\frac{(-3\,ht^2\,k + 3\,h\,ht\,kt - h\,(-htt\,k + h\,ktt))\,(hx\,k - h\,kx)^2}{h^9}\right], \{\}\right\}$$

```
solutionQ[syst2, J, xs, First@sol2]
```

True

```
xs = {h, ht, hx, k, kt, kx, λ, κ}; ys = J @@ xs;
eq1 = h * D[ys, ht] + k * D[ys, kt] == 0;
eq2 = h * D[ys, hx] + k * D[ys, kx] == 0;
eq3 = h * D[ys, h] + k * D[ys, k] + 2 * ht * D[ys, ht] +
    hx * D[ys, hx] + 2 * kt * D[ys, kt] + kx * D[ys, kx] -
    λ * D[ys, λ] == 0;
eq4 = h * D[ys, h] + k * D[ys, k] + ht * D[ys, ht] +
    2 * hx * D[ys, hx] + kt * D[ys, kt] + 2 * kx * D[ys, kx] -
    κ * D[ys, κ] == 0;
syst3 = {eq1, eq2, eq3, eq4};
sol3 = solveOverDetScalarFirstOrderPdes[syst3, J, xs] //
   Quiet
```

$$\left\{ J[h, ht, hx, k, kt, kx, \lambda, \kappa] \to C[4]\left[\frac{k}{h}, h\,\kappa\,\lambda, \frac{-ht\,k + h\,kt}{h^3\,\kappa}, \frac{(-hx\,k + h\,kx)\,\kappa}{h^2}\right], \{\} \right\}$$

```
solutionQ[syst3, J, xs, First@sol3]
```

True

### ◾ Test 2

The example treated below is a slight extension of the one found on the website [7].

```
xs = {t, x, y, z}; ys = f @@ xs;
eq1 = -y * D[ys, x] + z^2 * D[ys, z] + 3 * t * z * D[ys, t] -
    4 * z * ys - 3 * a * t^2 == 0;
eq2 = -y * D[ys, y] - z * D[ys, z] - t * D[ys, t] + ys == 0;
eq3 = -x * D[ys, y] - D[ys, z] == 0;
syst4 = {eq1, eq2, eq3};
sol4 = solveOverDetScalarFirstOrderPdes[syst4, f, xs] //
   Quiet
```

$$\left\{ f[t, x, y, z] \to -\frac{3\sqrt{2}\,a\,t^2\,x}{\sqrt{2\,y - 2\,x\,z}\,\sqrt{y - x\,z}} + \frac{t^{3/2}\,C[4]}{\sqrt{2\,y - 2\,x\,z}}, \{\} \right\}$$

```
solutionQ[syst4, f, xs, First@sol4]
```

True

### ▪ Test 3

We took the examples of this from the book of Saltykow [8].

```
xs = {y1, y2, y3, y4}; ys = f@@xs;
eq1 =
 D[ys, y1] - (1 / (y3 * y4)) * D[ys, y3] + (1 / y3^2) * D[ys, y4] ==
   0;
eq2 = D[ys, y2] + (1 / y4) * D[ys, y3] - (2 / y3) * D[ys, y4] == 0;
syst5 = {eq1, eq2};
sol5 = solveOverDetScalarFirstOrderPdes[syst5, f, xs] //
   Quiet
```

$\{f[y1, y2, y3, y4] \to C[2][y2 + y3\, y4,\ y1 + y3^2\, y4], \{\}\}$

```
solutionQ[syst5, f, xs, First@sol5]
```

True

```
xs = {y1, y2, y3, y4}; ys = f@@xs;
eq1 = 2 * y1 * D[ys, y1] + 3 * y2 * D[ys, y2] + 4 * y3 * D[ys, y3] +
    5 * y4 * D[ys, y4] == 0;
eq2 = D[ys, y1] + 4 * y1 * D[ys, y3] + 5 * y2 * D[ys, y4] == 0;
eq3 = y2 * D[ys, y3] + 2 * (y3 - 2 * y1^2) * D[ys, y4] == 0;
syst6 = {eq1, eq2, eq3};
sol6 =
 solveOverDetScalarFirstOrderPdes[syst6, f, xs] //
    Simplify // Quiet
```

$\left\{f[y1, y2, y3, y4] \to C[3]\left[\dfrac{-4\, y1^4 - 5\, y1\, y2^2 + 4\, y1^2\, y3 - y3^2 + y2\, y4}{y1\, y2^2 \left(\frac{y2}{y1^{3/2}}\right)^{2/3}}\right], \{\}\right\}$

```
solutionQ[syst6, f, xs, First@sol6]
```

True

### ■ Test 4

The systems of PDEs treated here are to be found in the book by Mansion [4].

```
xs = {y1, y2, y3, y4}; ys = v @@ xs;
eq1 = 2 * y2 * y4^2 * D[ys, y1] + y3^2 * y4 * D[ys, y4] - y3^2 * ys ==
   0;
eq2 = 2 * y2 * D[ys, y2] - y4 * D[ys, y4] - ys == 0;
eq3 = y2 * y4^2 * D[ys, y3] + y1 * y3 * y4 * D[ys, y4] -
   y1 * y3 * ys == 0;
syst7 = {eq1, eq2, eq3};
sol7 = solveOverDetScalarFirstOrderPdes[syst7, v, xs] //
   Quiet
```

$$\left\{ v[y1, y2, y3, y4] \to \sqrt{2} \sqrt{y2} \sqrt{y2\, y4^2}\; C[3]\left[\frac{1}{2}\left(-y1\, y3^2 + y2\, y4^2\right)\right], \{\}\right\}$$

```
solutionQ[syst7, v, xs, First@sol7]
```

True

```
xs = {y1, y2, y3, y4}; ys = z @@ xs;
eq1 =
 y1 * D[ys, y1] - y2 * D[ys, y2] + y3 * D[ys, y3] - y4 * D[ys, y4] ==
   0;
eq2 =
 y3 * D[ys, y1] + y4 * D[ys, y2] - y1 * D[ys, y3] -
    y2 * D[ys, y4] == 0; syst8 = {eq2, eq1};
sol8 = solveOverDetScalarFirstOrderPdes[syst8, z, xs] //
    Simplify // Quiet
```

$$\left\{z[y1, y2, y3, y4] \to C[2]\left[\frac{\sqrt{1 + \frac{y1^2}{y3^2}}\ y3\ (y1\ y2 + y3\ y4)}{\sqrt{2}\ \sqrt{y1^2 + y3^2}},\ \frac{\sqrt{1 + \frac{y1^2}{y3^2}}\ y3\ (y2\ y3 - y1\ y4)}{\sqrt{2}\ \sqrt{y1^2 + y3^2}}\right], \{\}\right\}$$

```
solutionQ[syst8, z, xs, First@sol8]
```

True

```
xs = {y1, y2, y3, y4, y5}; ys = f @@ xs;
eq1 = D[ys, y1] * D[ys, y5] - y2 * y4 == 0;
eq2 = D[ys, y2] * D[ys, y4] - y1 * y5 == 0;
syst9 = {eq1, eq2};
sol9 =
 solveOverDetScalarFirstOrderPdes[syst9, f, xs] //
    Simplify // Quiet
```

$$\left\{ f[y1, y2, y3, y4, y5] \to \right.$$
$$C[1][y3, y4, y5] + \frac{y1\, y4\, y5}{C[2][y3, y4, y5]} + y2\, C[2][y3, y4, y5],$$
$$\left\{ \left( x119\, x120 \left( \frac{x119\, y1}{C[2][x118, x119, x120]} + C[1]^{(0,0,1)}[x118, x119, x120] + y2\, C[2]^{(0,0,1)}[x118, x119, x120] - \frac{x119\, x120\, y1\, C[2]^{(0,0,1)}[x118, x119, x120]}{C[2][x118, x119, x120]^2} \right) \right) \right/$$
$$C[2][x118, x119, x120] == x119\, y2,$$
$$\left( -x119\, x120\, y1\, C[2]^{(0,1,0)}[x118, x119, x120] + C[2][x118, x119, x120]^2 \left( C[1]^{(0,1,0)}[x118, x119, x120] + y2\, C[2]^{(0,1,0)}[x118, x119, x120] \right) \right) /$$
$$\left. \left. C[2][x118, x119, x120] == 0 \right\} \right\}$$

The second entry of **sol9** indicates that there are two PDEs that were not solved. These PDEs can be straightforwardly separated with respect to **y1** and **y2** to obtain new PDEs that are easily solved using the built-in function **DSolve**.

### ▪ Test 5

The last example of this subsection is due to Boole [ ].

```
xs = {x, y, z, t}; ys = f @@ xs;
eq1 = D[ys, t] + (1 - x) * D[ys, x] / t == 0;
eq2 = D[ys, z] + (y - (x - 1) * t) * D[ys, x] / (z * t) == 0;
eq3 = D[ys, y] + D[ys, x] / t == 0; syst10 = {eq1, eq2, eq3};
sol10 = solveOverDetScalarFirstOrderPdes[syst10, f, xs] //
   Quiet
```

{f[x, y, z, t] → C[3][(-t (-1 + x) + y) z], {}}

```
solutionQ[syst10, f, xs, First@sol10]
```

True

```
xs = {x, y, z}; ys = u @@ xs;
eq1 = D[ys, x] == 4 * Sin[y] * Sin[y] * Cos[z];
eq2 = (1 / x) * D[ys, y] == 4 * Cos[z] * Sin[2 * y];
eq3 = (1 / (x * Sin[y])) * D[ys, z] == -4 * Sin[y] * Sin[z];
syst11 = {eq1, eq2, eq3};
sol11 = solveOverDetScalarFirstOrderPdes[syst11, u, xs] //
   Quiet
```

{u[x, y, z] → C[3] + x (-Cos[2 y - z] + 2 Cos[z] - Cos[2 y + z]), {}}

```
solutionQ[syst11, u, xs, First@sol11]
```

True

```
xs = {x, y}; ys = z @@ xs; eq1 = D[ys, x] == a * E^(y - ys);
eq2 = D[ys, y] == b * E^(y - ys) + 1; syst12 = {eq1, eq2};
sol12 = solveCompatiblePdes[syst12, z, xs] // Quiet
```

{z[x, y] → Log[a e^y x + b e^y y + e^y C[2]], {}}

```
solutionQ[syst12, z, xs, First@sol12]
```

True

```
xs = {x, y, z, t}; ys = p @@ xs;
eq1 =
 D[ys, x] + (t + x*y + x*z)*D[ys, z] + (y + z - 3*x)*D[ys, t] ==
   0;
eq2 = D[ys, y] + (x*z*t + y - x*y)*D[ys, z] +
    (z*t - y)*D[ys, t] == 0;
syst13 = {eq1, eq2};
sol13 =
 solveOverDetScalarFirstOrderPdes[syst13, p, xs] //
    Simplify // Quiet
```

$$\left\{p[x, y, z, t] \to C[3]\left[-t\,x - x^3 - \frac{y^2}{2} + z\right], \{\}\right\}$$

```
solutionQ[syst13, p, xs, First@sol13]
```

True

# ■ Conclusion

In this work we have introduced and implemented an algorithm based on Bour-Mayer method, for solving an overdetermined systems of PDEs in one unknown. We have demonstrated the efficiency of our approach through the consideration of several examples.

# ■ Acknowledgments

I gratefully acknowledge partial financial support from the DST-NRF Centre of Excellence in Mathematical and Statistical Sciences, School of Computer Science and Applied Mathematics, University of theWitwatersrand, Johannesburg, South Africa. I thank Prof. F. M. Mahomed for securing the necessary funds and his team for the hospitality during my visit last summer. This work is dedicated to my daughter Katlego on her sixteenth birthday.

# ■ References

## About the Author

Dr. C. Wafo Soh is currently an Associate Professor of Mathematics at Jackson State University and a Visiting Associate Professor of Applied Mathematics at the University of Witwatersrand. He is the co-founder of the South African start-up " Recursive Thinking Consulting" which specializes in process mining.

**Célestin Wafo Soh**
*Department of Mathematics andStatistical Science*
*JSU Box 1760, Jackson State University*
*1400 JR Lynch Street*
*Jackson,MS 39217*
*Celestin.Wafo_Soh@jsums.edu*